\documentclass[twocolumn]{aa} 
\usepackage{graphicx}
\usepackage{amsmath}
\usepackage{txfonts}
\usepackage{lscape}
\begin{document}
\newcommand{\bvec}[1]{\mbox{\boldmath $#1$}}

   \title{CH radio emission from heiles cloud 2 as a tracer of molecular cloud evolution}

   \subtitle{}

   \author{N. Sakai
          \inst{1}, H. Maezawa\inst{2}, T. Sakai\inst{3}, K. M. Menten\inst{4}, 
          \and
          S. Yamamoto\inst{1}
          }

   \institute{Department of Physics and Research Center for the Early Universe, The University of Tokyo, Bunkyo-ku, Tokyo 113-0033, Japan
              \email{nami@taurus.phys.s.u-tokyo.ac.jp} 
              \and
              Physical Science, Osaka Prefecture University, Gakuen-cho, Naka-ku, Sakai-shi, 599-8531, Japan
              \and
              Institute of Astronomy, The University of Tokyo, Osawa, Mitaka, Tokyo, 181-8588, Japan
              \and
              Max-Planck-Institut fuer Radioastronomie, Auf dem H$\ddot{\rm u}$egel 69, D-53121 Bonn, Germany
              }
              
   \date{Received 24 February 2012; Accepted 28 August 2012}

 
  \abstract
   {}
   {A mapping observation of the $J=1/2$ $\Lambda$-type doubling transition (3.3 GHz) of CH has been conducted toward Heiles Cloud 2 (HCL2) in the Taurus molecular cloud complex to reveal its molecular cloud-scale distribution.}
   {The observations were carried out with the Effelsberg 100~m telescope.}
   {The CH emission is found to be extended over the whole region of HCL2.  It is brighter in the southeastern part, which encloses the TMC-1 cyanopolyyne peak than in the northwestern part.  Its distribution extends continuously from the peak of the neutral carbon emission (CI peak) to the TMC-1 ridge, as if it were connecting the distributions of the [C I] and C$^{18}$O emissions.  Since CH is an intermediate in gas-phase chemical reactions from C to CO, its emission should trace the transition region.  The above distribution of the CH emission is consistent with this chemical behavior.  Since the CH abundance is subject to the chemical evolutionary effect, the CH column density in HCL2 no longer follows a linear correlation wit the H$_2$ column density reported for diffuse and translucent clouds.  More importantly, the CH line profile is found to be composed of the narrow and broad components.  Although the broad component is dominant around the CI peak, the narrow component appears in the TMC-1 ridge and dense core regions such as L1527 and TMC-1A.  This trend seems to reflect a narrowing of the line width during the formation of dense cores.  These results suggest that the 3.3 GHz CH line is a useful tool for tracing the chemical and physical evolution of molecular clouds. 
}
   {}

   \titlerunning{CH in HCL2}
   \maketitle
%

\section{Introduction}

It is widely recognized that chemical abundances show significant variations within a molecular cloud as well as between one cloud and another.  Understanding this differentiation is an important topic of astrochemistry and astrophysics.  A most famous and well-studied case is the TMC-1 ridge, which is a part of Heiles Cloud~2 (HCL2) in the Taurus molecular cloud complex.  Carbon-chain molecules such as CCS and HC$_3$N are abundant in its southeastern part ($^{\prime}$cyanopolyyne peak$^{\prime}$; referred to hereafter as TMC-1(CP)), whereas NH$_3$ and N$_2$H$^+$ are abundant in the northwestern part ($^{\prime}$ammonia peak$^{\prime}$; referred to hereafter as TMC-1(NH$_3$)) (e.g. \cite{lit79, hir92, ola88, pra97}).  This chemical differentiation is most naturally interpreted in terms of an evolutionary effect (e.g. \cite{hir92}), although other possibilities have also been proposed (e.g. \cite{mar00, gar07}).  Carbon-chain molecules are generally abundant in early stages of the chemical evolution before C is not completely locked up in CO, while they become deficient in more advanced stages due to gas-phase destruction and depletion onto dust grains (\cite{suz92, ben98, aik03}).  In contrast, NH$_3$ and N$_2$H$^+$ appear in advanced stages because of their slow formation (\cite{suz92, ben98}).  In fact, the protostar IRAS~04381+2540 has been found near the ammonia peak of the TMC-1 ridge (\cite{ter89}), testifying that this peak is, indeed, more chemically evolved than the cyanopolyyne peak.  All these results suggest that along the TMC-1 ridge we are observing an evolutionary trend from less advanced in its southeastern part to more advanced in the northwestern part.

This evolutionary scenario is supported by observations of the 492 GHz atomic carbon [C I] $^3P_1-^3P_0$ fine structure line with the Mount Fuji submillimeter-wave telescope (\cite{mae99}).  These authors conducted a large-scale survey of the [C I] line covering the whole region of HCL2, and found that the [C I] emission is more intense in the southeastern part of HCL2 than in the northwestern part by a factor of 2.  On the other hand, the C$^{18}$O line is generally intense in the northwestern part.  Because the major form of carbon in the gas phase changes from C to CO, the direction of chemical evolution from less to more developed is again from the southeast to the northwest.  This is nicely consistent with the CCS and NH$_3$ observations of the TMC-1 ridge, and important for our understanding of the formation of molecular cloud cores.  A similar behavior of C, CO, and CCS on a much larger scale is observed in the AFGL333 region of the W3 giant molecular cloud by Sakai et al. (2006).

To examine the above formation process of HCL2 in more detail, we focus on observations of the CH emission (3.3 GHz).  CH is the first and simplest neutral molecule formed from C and/or C$^+$ in the gas phase and an important chemical intermediate in the production pathway of CO from C and C$^+$.  It is also suggested to be a key reactant in the production of carbon-chain molecules at early evolutionary stages (\cite{sak07a}).  Observations of the CH line toward the HCL2 region were first reported by Rydbeck et al. (1976), although only toward two positions in HCL2 were observed with the 15$^{\prime}$ beam.  Very recently, Suutarinen et al. (2011) reported observations of the CH line along the TMC-1 with the MPIfR 100~m telescope.  They found that the abundance of CH has a peak toward TMC-1(CP) on the basis of a strip map along the ridge.  However, these authors only focused on the TMC-1 ridge, leaving the overall distribution of CH in HCL2 unexplored.

The CH emission (3.3~GHz) has extensively been observed for diffuse clouds, translucent clouds, bright limbed clouds, outflows, and dark clouds (e.g. \cite{lan78, san80, san81, mat86, jac87, san88, mag92, mag93b}).  Many of these previous efforts focused on the correlation between the CH column density and the optical extinction.  With the aid of the optical data, the CH column density is recognized as a good tracer of the H$_2$ column density in diffuse and translucent clouds (e.g. \cite{fed82, dan84, mat86, she08}).  Recently, Chastain et al. (2010) conducted the high spatial resolution observation of the high-latitude translucent clouds MBM~3 and MBM~40 with the NAIC~305~m telescope at Arecibo, and found a slight difference between the distribution of CH and that of CO.  This may suggest that the CH column density does not always correlate to the H$_2$ column density, and that we have to consider chemical evolutionary effects on the CH abundance.  With these in mind, we have explored the distribution of the CH line in a whole region of HCL2.


\section{Observations}

Observations were carried out with the MPIfR 100 m telescope at Effelsberg in August 2007, September 2008, and September 2009.  We observed two hyperfine components of the $J=1/2$ $\Lambda$-type doubling transition of CH ($F=1-1$; $\nu=3.335482(1)$ GHz; $S\mu ^{2}=1.4205$ Debye$^{2}$: $F=0-1$; $\nu=3.349193(3)$ GHz; $S\mu ^{2}=0.7105$ Debye$^{2}$) (\cite{mcc06, mul05} (CDMS)) toward HCL2.  Here, the numbers in parentheses represent the 1~$\sigma$ error in units of the last significant digits.  We did not observe the $F=1-0$ line ($\nu=3.263795(3)$ GHz) because of the limitation of the backend.  We used the 9 cm receiver as a frontend, whose system noise temperature was about 35~K.  The beam size and the main beam efficiency of the telescope are 3$^{\prime}$.8 and 0.73, respectively.  The telescope pointing was checked every three hours by observing nearby continuum sources, and was maintained to be better than 20$^{\prime\prime}$.  A small daily variation of the intensity was calibrated by repeatedly observing the $F=1-1$ main line toward TMC-1(CP).  Our backend was the fast fourier transform spectrometer, FFTS, with a bandwidth and resolution of 20~MHz and 1.2~kHz, respectively.  This frequency resolution corresponds to a velocity resolution of 0.1~km~s$^{-1}$.   The observations were carried out in the frequency-switching mode with a frequency offset of 0.2~MHz.  The whole HCL2 region was mapped in the $F=1-1$ main line.  Toward a few selected positions, we observed the $F=1-1$ main line and the $F=0-1$ satellite line with a higher signal to noise (S/N) ratio to investigate the line profile.


\section{Results}
\subsection{Line profile}
\begin{figure}
   \centering
      \includegraphics[width=9cm]{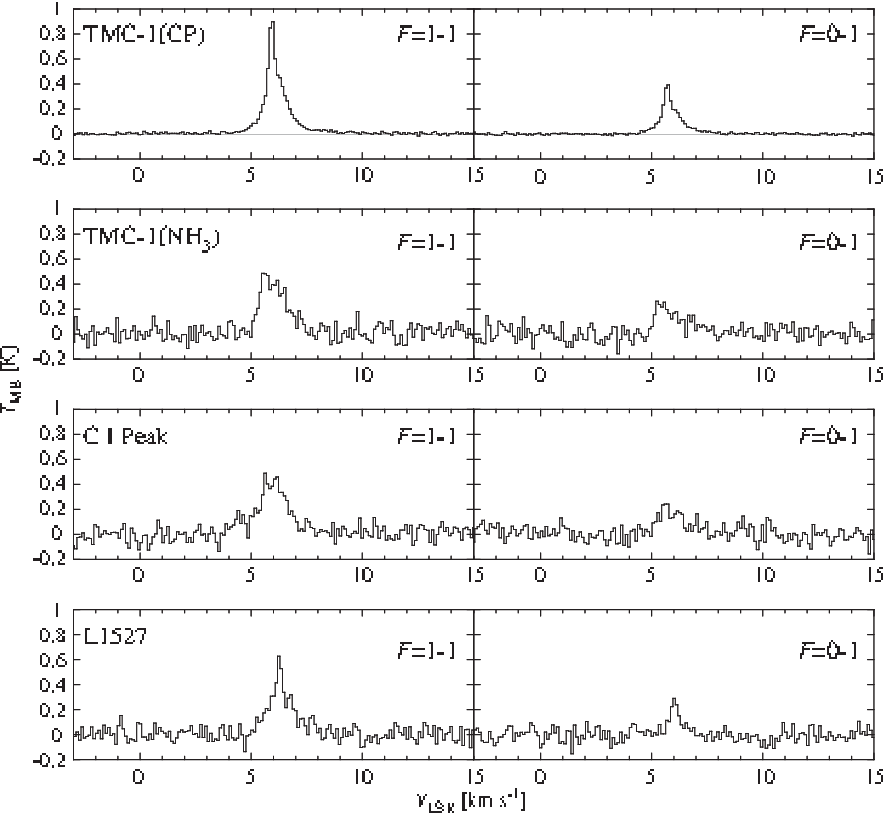}
      \caption{Line profiles of CH toward four representative positions in HCL2. }
         \label{fig:line}
\end{figure}

\begin{figure}
   \centering
      \includegraphics[width=5cm]{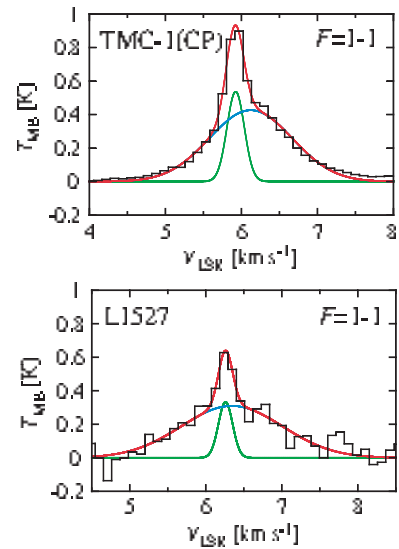}
      \caption{Examples of the double Gaussian fit for TMC-1(CP) and L1527.}
         \label{fig:fit}
   \end{figure}

Our first CH observation toward TMC-1(CP) was conducted in 2007.  After a long on-source integration of 25 hours for a search for $^{13}$CH, the CH lines were detected with a high S/N ratio, as shown in Figure \ref{fig:line}.  In addition to the main line ($F = 1-1$), the satellite line ($F = 0-1$) was also detected.  The intensity ratio between the main and satellite lines is close to the intrinsic intensity ratio of 2, indicating that the lines are optically thin.  As seen in Figure \ref{fig:line}, the line profile apparently consists of narrow and broad components both for the main and the satellite lines.  The two components can be separated by fitting two Gaussian profiles, as shown in Figure \ref{fig:fit}.  The derived parameters are given in Table \ref{tab:para}.

\begin{table*}
\caption{Line parameters of the CH emission observed toward several cores.\label{tab:para}}
\label{table:2}
\centering
\begin{tabular}{llcccc}
\hline\hline
Source&Transition&$T_{\rm MB}$&$\Delta V$&$V_{\rm LSR}$&$\int T_{\rm MB} dv$\\
$(\alpha_{2000}, \delta_{2000})$&$\quad$&[K]&[km~s$^{-1}$]&[km~s$^{-1}$]&[K~km~s$^{-1}$]\\
\hline
TMC-1(CP)&$F=1-1$ narrow&0.536(12)&0.26(1)&5.916(3)&0.787(3)\\
{\small $(04^{\rm h} 41^{\rm m} 42^{\rm s}.88, 25^{\circ} 41^{\prime} 27^{\prime\prime}.0)$}&$F=1-1$ broad&0.429(8)&1.28(2)&6.123(8)&--\\
$\quad$&$F=0-1$ narrow&0.237(8)&0.27(1)&5.693(4)&0.333(3)\\
$\quad$&$F=0-1$ broad&0.181(5)&1.29(3)&5.892(13)&--\\
TMC-1(NH$_3$)&$F=1-1$&0.445(22)&1.40(8)&5.97(3)&0.665(22)\\
{\small $(04^{\rm h} 41^{\rm m} 18^{\rm s}.50, 25^{\circ} 48^{\prime} 13^{\prime\prime}.6)$}&$F=0-1$&0.205(21)&1.44(17)&5.81(7)&0.349(21)\\
CI peak&$F=1-1$&0.407(21)&1.60(9)&5.97(4)&0.759(23)\\
{\small $(04^{\rm h} 42^{\rm m}08^{\rm s}.70, 25^{\circ} 21^{\prime} 25^{\prime\prime}.3)$}&$F=0-1$&0.193(20)&1.46(18)&5.85(8)&0.305(22)\\
L1527&$F=1-1$ narrow&0.333(58)&0.22(5)&6.26(2)&0.607(22)\\
{\small $(04^{\rm h} 39^{\rm m} 53^{\rm s}.89, 26^{\circ} 03^{\prime} 11^{\prime\prime}.0)$}&$F=1-1$ broad&0.310(30)&1.57(14)&6.34(5)&--\\
$\quad$&$F=0-1$ narrow&0.204(64)&0.31(11)&6.02(3)&0.179(17)\\
$\quad$&$F=0-1$ broad&0.089(57)&1.13(46)&6.12(16)&--\\
\hline
\end{tabular}
\begin{list}{}{}
\item[Note.] $T_{\rm MB}$, $\Delta V$, and$V_{\rm LSR}$ are obtained by a Gaussian fit.  The error in the parentheses denotes one standard deviation in units of the last significant digits.  The $V_{\rm LSR}$ values refer to the rest frequencies given in Section~2, and the error of $V_{\rm LSR}$ does not include the uncertainty of the rest frequency.  A slight difference of the $V_{\rm LSR}$ values between the $F=1-1$ (main) and $F=0-1$ (satellite) component likely originates from the error of the rest frequency of the $F=0-1$ component.  See text for details.
\end{list}
\end{table*}

The narrow component of the main line has an LSR velocity, $V_{\rm LSR}$, of 5.916(3)~km~s$^{-1}$ and a line width, $\Delta V$, of 0.26(1)~km~s$^{-1}$, where the numbers in parentheses represent the fitting error in units of the last significant digits.  This $V_{\rm LSR}$ value is close to the velocities observed in various molecular lines toward this position (e.g. \cite{dic01, kai04, sak08}), whereas the line is slightly narrower than those of other molecules.  It should be noted that the 1 $\sigma$ uncertainty of the rest frequencies for the main and satellite hyperfine components are 1 and 3 kHz, respectively, corresponding to velocities of 0.09 and 0.27 km s$^{-1}$, respectively (Table \ref{tab:para}) (\cite{mcc06}).  Therefore, we cannot definitively identify which velocity component appears as the narrow component among several velocity components known for other molecules (e.g. \cite{lan95}).  We also note that the $V_{\rm LSR}$ value of the narrow component for the satellite line is lower by 0.2~km~s$^{-1}$ than that of the main line.  This likely also originates from the insufficient accuracy of the rest frequency measurement.

In contrast to the narrow component, the broad component has so far not been discussed broadly toward TMC-1(CP).  Motivated by this result, we additionally observed the CH emission toward three other positions in HCL2, TMC-1(NH$_3$), L1527, and the CI peak, where the CI peak is the intensity peak of the [C I] emission in HCL2 (\cite{mae99}).  The CH lines were detected toward all positions, as shown in Figure \ref{fig:line}.  Again, the intensity ratios between the main and satellite lines are close to 2.  On the other hand, the line shape is found to vary from position to position.  The broad and narrow components are also seen in L1527, whereas only the broad component is seen in TMC-1(NH$_3$) and the CI peak.  We fitted the observed profile for L1527 by two Gaussian functions (Figure \ref{fig:fit}), while we employed single Gaussian fitting for TMC-1(NH$_3$) and the CI peak.  The line parameters are summarized in Table \ref{tab:para}.

\subsection{Column densities}
The column density of CH is evaluated by assuming optically thin emission.  This is justified, because the intensity ratio between the main and satellite hyperfine components is close to the ratio of their intrinsic intensities, as mentioned before.  HCL2 is a quiescent cloud not influenced by high-mass star formation activities, although several low-mass protostars such as IRAS~04381+2540 and IRAS~04368+2557 are associated with it.  Therefore, the gas kinetic temperature is as low as 10 K in its dense parts.  This means that almost all the CH molecules are populated in the $J=1/2$ $\Lambda$-type doubling levels, and the populations of the higher $J$ levels are negligible.  In the cloud peripheries, the temperature would be higher than 10 K due to photoelectric heating by interstellar UV radiation.  In this case, the density is too low to excite CH even to the first rotationally excited state ($J=3/2$), because the critical density required for the excitation of the lowest rotational transition ($J=3/2-1/2$; 530 GHz) is as high as $10^6$ cm$^{-3}$.   Hence, practically all CH molecules are in the $J=1/2$ levels.  For this reason, we only considered the $J=1/2$ levels for calculations of the partition function in the evaluation of the column density.

\begin{table}
\caption{Column densities of CH.\label{tab:column}}
\label{table:3}
\centering
\begin{tabular}{llc}
\hline\hline
Sources&$\quad$&$N$(CH) [cm$^{-2}$] \\
\hline
TMC-1(CP)&narrow&$(0.41\pm 0.05)\times 10^{14}$\\
$\quad$&broad&$(1.57\pm 0.12)\times 10^{14}$\\
TMC-1(NH$_3$)&$\quad$&$(1.79\pm 0.42)\times 10^{14}$\\
CI peak&$\quad$&$(1.88\pm 0.45)\times 10^{14}$\\
L1527&narrow&$(0.21\pm 0.18)\times 10^{14}$\\
$\quad$&broad&$(1.39\pm 0.57)\times 10^{14}$\\
\hline
\end{tabular}
\begin{list}{}{}
\item[Note.] The error is evaluated from three times the standard deviation of the line parameters.  $T_{ex}$ is assumed to be $-60$~K. See text.
\end{list}
\end{table}

It is well known that the populations of the $J=1/2$ $\Lambda$-type doubling levels are inverted over a wide range of physical conditions (\cite{gen79, buj84, san88, lis02}).  Hence, the excitation temperatures are usually negative for the $F=1-1$, $1-0$, and $0-1$ lines.  In particular, a strong excitation anomaly occurs for the $F=0-1$ line, which sometimes appears as an absorption against the background continuum sources (e.g. \cite{gen79}).  In contrast, the $F=1-1$ main line behaves quite normally.  Under the assumption of optically thin emission, the column density of CH, $N$(CH), can be related to the integrated intensity ($W$ in K km~s$^{-1}$) of the $F=1-1$ line as
$$N({\rm CH})=2.82 \times 10^{14} \ T_{\rm ex}/(T_{\rm ex}-T_{\rm b}) \ W \ {\rm cm}^{-2}, \eqno(1)$$
\noindent
where $T_{\rm ex}$ and $T_{\rm b}$ represent the excitation temperature and the cosmic microwave background temperature, respectively (e.g. \cite{san88}).  Assuming an effective excitation temperature of $-60$~K reported for dark clouds (\cite{gen79}), we calculated the column densities, as summarized in Table \ref{tab:column}.  As expected from eq.(1), the uncertainty of the excitation temperature does not contribute to the error of the column density very much, unless $T_{\rm ex}$ is as low as  $T_{\rm b}$.  Even if we change the excitation temperature by $\pm30$~K, the column density varies only within 5~\%.

\begin{figure*}
   \centering
      \includegraphics[width=17cm]{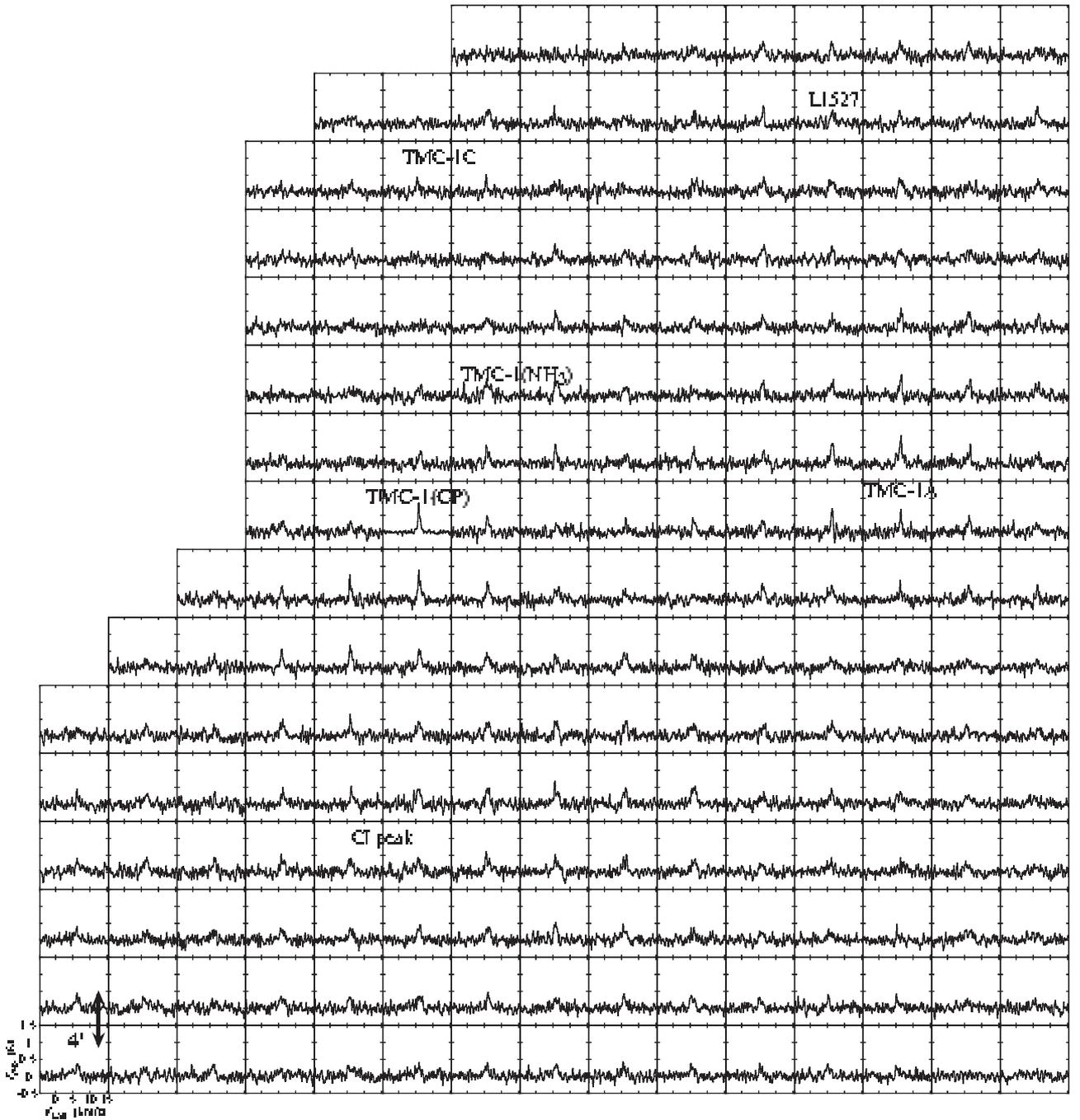}
      \caption{Profile map of the CH ($F=1-1$) main line.  The spacing of the spectra is 4$^{\prime}$ in each direction.  Approximate positions of the CI peak, TMC-1(CP), TMC-1(NH$_3$), L1527, TMC-1C, and TMC-1A are given.  The typical rms noise level is about 0.1~K.}
         \label{fig:chpmap}
\end{figure*}

\subsection{Distribution}
\begin{figure*}
   \centering
      \includegraphics[width=17cm]{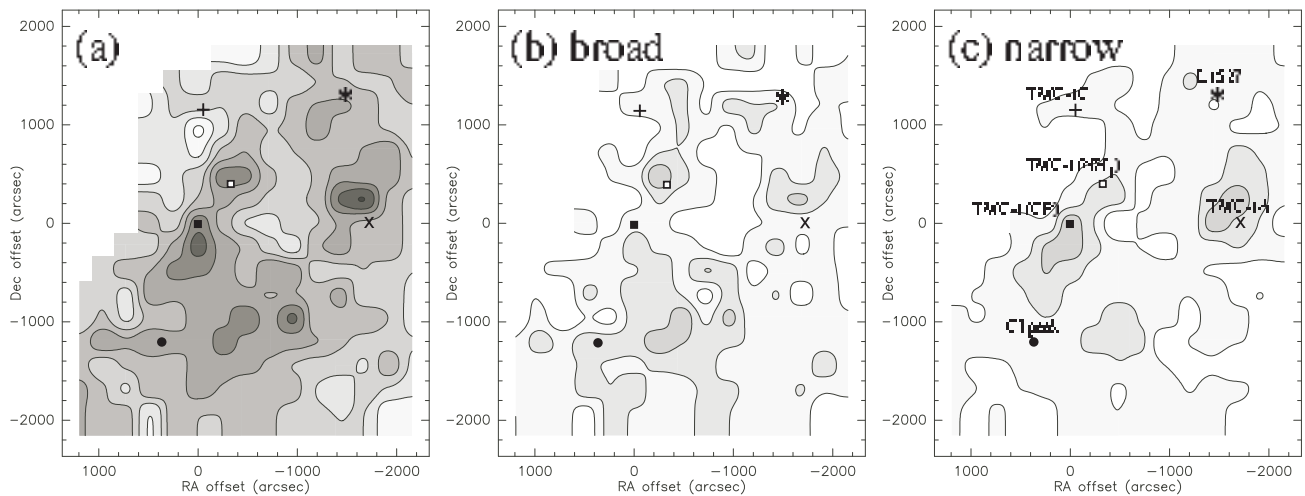}
      \caption{(a) Integrated intensity map of the CH emission from  5.0 to 7.2~km~s$^{-1}$.  Contours show every 0.1~K~km~s$^{-1}$ from 0.1~K~km~s$^{-1}$.  (b) Integrated intensity map of the $broad$ component of the CH emission (4.7-5.8 and 6.6-7.2~km~s$^{-1}$).  Contours show every 0.1~K~km~s$^{-1}$ from 0.2~K~km~s$^{-1}$.  (c) Integrated intensity map of the $narrow$ component of the CH emission.  Since the narrow component velocity slightly changes in a large scale, the 0.2~km~s$^{-1}$ span centered at the peak velocity is integrated to extract the narrow component.  Hence, it should be noted that a part of the broad component is contaminated.  Contours show every 0.05~K~km~s$^{-1}$ from 0.1~K~km~s$^{-1}$.  Filled circle, filled square, open square, star, cross mark,  and x symbols correspond to the position of the CI peak, TMC-1(CP), TMC-1(NH$_3$), L1527, TMC-1C, and TMC-1A, respectively, as shown in (c).}
         \label{fig:intmap2}
\end{figure*}

\begin{figure*}
   \centering
      \includegraphics[width=15cm]{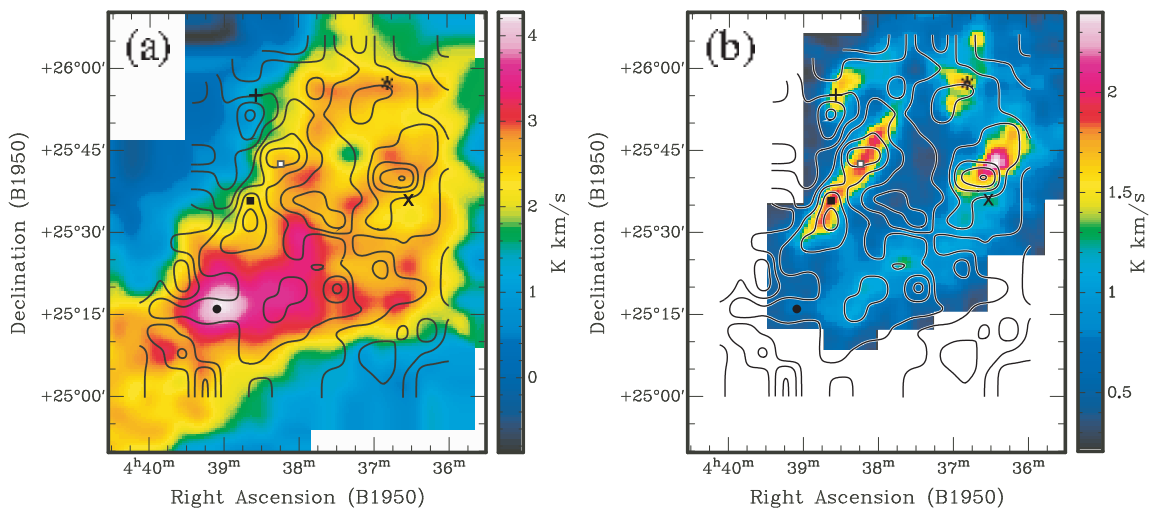}
      \caption{(a) Contours show the integrated intensity map of CH overlaid on the color image of the [C I] ($^3P_1-^3P_0$: 492~GHz) distribution taken with the Mount Fuji submillimter telescope (\cite{mae99}).  The velocity range of the integration for the CH line is from 5.0 to 7.2~km~s$^{-1}$.  The lowest contour and the contour interval is 0.1~K~km s$^{-1}$. (b)  Contours show the integrated intensity map of CH overlaid on the color image of the C$^{18}$O ($J=1-0$) distribution taken with the Nobeyama 45 m telescope (archival data).  The velocity range of the integration of the C$^{18}$O line is from 5.0 to 7.2~km~s$^{-1}$.}
         \label{fig:intmap}
\end{figure*}

To explore the distribution of the CH emission and the origin of the line shape variation, we then conducted mapping observations of the whole HCL2 region.  To do this within the limited observation time, we employed a roughly FWHM-sized grid spacing of 4$^{\prime}$. Therefore, the map is undersampled.  The integration time for each position was typically 10 min.  Figure \ref{fig:chpmap} shows a profile map of the CH main line, and Figure \ref{fig:intmap2}a shows the corresponding integrated intensity map.  The CH emission is distributed throughout the HCL2 region.  The ring structure of HCL2 can be seen, and the TMC-1 ridge structure is also evident.  In addition to these, the CH emission is observed with moderate intensities around the CI peak.  It is generally brighter in the southeastern part, which includes TMC-1(CP), than in the northern part.  Figures \ref{fig:intmap}a and \ref{fig:intmap}b show the integrated intensity map of CH overlaid on the images of the [C I] and C$^{18}$O ($J=1-0$) emission, respectively.  The distribution of CH looks continuous from the CI peak to the TMC-1 ridge, as if it connected the distributions of [C I] and C$^{18}$O.

Next, we focused on the line profile.  As shown in Figure \ref{fig:chpmap}, the line profile systematically changes from the CI peak to the TMC-1 ridge.  The CH line is quite broad (typically 3~km~s$^{-1}$) toward the CI peak and its surroundings, while a narrower component appears in the TMC-1 ridge traced by the C$^{18}$O and CCS lines.  Both the broad and the narrow components are blended toward the cyanopolyyne peak, as mentioned before.  When we look at the TMC-1C core, which is located north of the TMC-1 ridge i$\Delta \alpha$= -1$^{\prime}$, $\Delta \delta $= 19$^{\prime}$ from TMC-1(CP)), a narrow component is dominant.  A similar line-profile change can be seen from the south part of HCL2 to the TMC-1A region.

The maps of the narrow and broad components are shown in Figures \ref{fig:intmap2}b and \ref{fig:intmap2}c, respectively.  The narrow component seems to be associated with the dense cores in HCL2, such as the TMC-1 ridge and TMC-1A.  Note that there exists a small region in the southwest end of HCL2 i$\Delta \alpha$= -28$^{\prime}$, $\Delta \delta $= -24$^{\prime}$ from TMC-1(CP)) that exhibits a narrow line profile (See Figure \ref{fig:chpmap}).  There is no report of a dense core toward this direction, as far as we know, and hence, its origin is unclear.  On the other hand, the broad component is extended over the HCL2 cloud, even toward the southern part around the CI peak (Figure \ref{fig:intmap2}b), where the H$_2$ density ($5.4 \times 10^3$~cm$^{-3}$) is lower than that in the dense cores ($\sim 10^5$~cm$^{-3}$) (\cite{mae99}).  Judging from the wide distribution of the broad component, it likely traces a less-dense envelope surrounding the dense cores of HCL2.

\section{Discussion}
\subsection{Abundance of CH}

The column densities of CH toward TMC-1(CP) are evaluated to be $(0.41\pm 0.05)\times 10^{14}$ and $(1.57\pm 0.12)\times 10^{14}$ cm$^{-2}$ for the narrow and broad components, respectively.  The error denotes three times the standard deviation.  The total column density is $(1.98\pm 0.17)\times 10^{14}$ cm$^{-2}$.  Rydbeck et al. (1976) reported the column density of CH toward TMC-1 to be $3 \times 10^{14}$~cm$^{-2}$ with the 15$^{\prime}$ beam, assuming a beam dilution factor of 2.  The total column density derived in our study is consistent with theirs.  Furthermore, our column density toward TMC-1(CP) is also consistent with that reported by Suutarinen et al. (2011).

If we assume an H$_2$ column density of $1.5 \times 10^{22}$~cm$^{-2}$ for the narrow component (\cite{mae99}), we calculate a fractional abundance of CH of $2.7 \times 10^{-9}$.   The H$_2$ column density of the broad component is uncertain.  Obviously, its upper limit would be the H$_2$ column density of the narrow component.  Then, the fractional abundance of the broad component is higher than $1.0 \times 10^{-8}$.  This indicates that the CH abundance is higher in the less dense component.  In dense core regions, the reaction of CH with O proceeds rapidly to produce CO, and hence, the CH abundance becomes relatively low, particularly in the late stages of chemical evolution (e.g. \cite{her89}).  A similar trend can also be seen for the column densities of the narrow and broad components in L1527, as shown in Table \ref{tab:column}.  Note that the narrow component emission may be diluted by the telescope beam (3$^{\prime}$.8), since the dense core of L1527 is as small as 2$^{\prime}$ (\cite{ben89}).

On the other hand, the CH column density toward the CI peak is evaluated to be $(1.88\pm 0.45)\times 10^{14}$ cm$^{-2}$.  Since the H$_2$ column density toward this position is reported to be $6 \times 10^{21}$~cm$^{-2}$ or higher (\cite{mae99}), the fractional abundance of CH is evaluated to be about $3 \times 10^{-8}$.  The column density and the fractional abundance of CH are close to those found in the broad component toward TMC-1(CP).  Furthermore, they are almost comparable to the CH abundance in diffuse clouds ($(4.3\pm 1.9)\times 10^{-8}$; \cite{lis02}: $(3.5^{+2.1}_{-1.4})\times 10^{-8}$; \cite{she08}).  The high abundance of CH is consistent with our interpretation that the broad component traces less-dense envelopes.

\subsection{Distribution of CH}
In the present study, we mapped the $0.8 \times 1.0$ deg$^2$ region of HCL2 in the $J=1/2$ $\Lambda$-type doubling transition of CH.  We here compare the distribution of CH with those of C$^{18}$O ($J=1-0$) and [C I].  As seen in Figure \ref{fig:intmap}b, the C$^{18}$O distribution traces the TMC-1 ridge well and also several low-mass star-forming cores such as L1527 and TMC-1A.  On the other hand, the [C I] emission shows a much more extended distribution (Figure \ref{fig:intmap}a).  In particular, the [C I] emission is extended to the southern part of HCL2, where the C$^{18}$O emission is fairly faint.  Hence, the distributions of C$^{18}$O and [C I] are anticorrelated to each other, as suggested by Maezawa et al. (1999).   They suggest that the anticorrelation represents a chemical evolutionary sequence from C to CO.  However, the [C I] and C$^{18}$O distributions are very different, and their mutual relationship is not very clear.  When we focus on the TMC-1 ridge traced by C$^{18}$O, for instance, its distribution is suddenly disrupted at its southern end, and there seems to exist a certain gap between the ridge and the CI peak.  Now, we find that the CH emission is distributed as if it continuously bridged the CI peak and the TMC-1 ridge.  Therefore, it is likely that the CI peak and the TMC-1 ridge are physically related to each other.

\begin{figure}
   \centering
      \includegraphics[width=8.5cm]{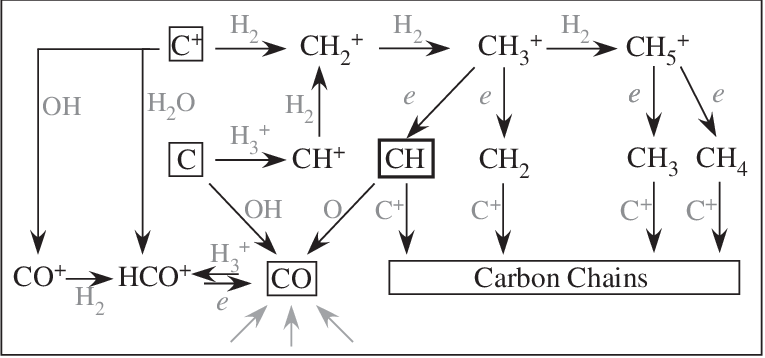}
      \caption{Schematic diagram of the basic hydrocarbon chemistry and its link to CO.}
         \label{fig:chem}
\end{figure}

This argument can be supported from the chemical point of view.  Figure \ref{fig:chem} shows a schematic diagram of the basic hydrocarbon chemistry and its connection to the production of CO.  CH is one of the simplest neutral species first produced from C$^+$ or C in molecular clouds.  The reaction of CH with O is an important route to produce CO, although there are other pathways producing CO, such as C + OH.  Hence, CH can be regarded as an intermediate species between C and CO.  The distribution of CH bridging the spatial gap between those of C$^{18}$O and [C I] seems to be a natural consequence of this chemical characteristic of CH, giving a yet another verification of the evolutionary picture in HCL2.

Furthermore, the distribution of the narrow component of CH in the TMC-1 ridge has a maximum at the cyanopolyyne peak.  The narrow component becomes weaker in the northern part of the ridge around the ammonia peak, just as in the case of carbon-chain molecules (e.g. \cite{lit79, hir92, ola88}).  Since CH is converted to CO rapidly in dense cores (e.g. \cite{her89}), it is abundant in the young stage.  This is consistent with the chemical youth of TMC-1(CP) in comparison with TMC-1(NH$_3$) (\cite{hir92}).

\begin{figure}
   \centering
      \includegraphics[width=8.5cm]{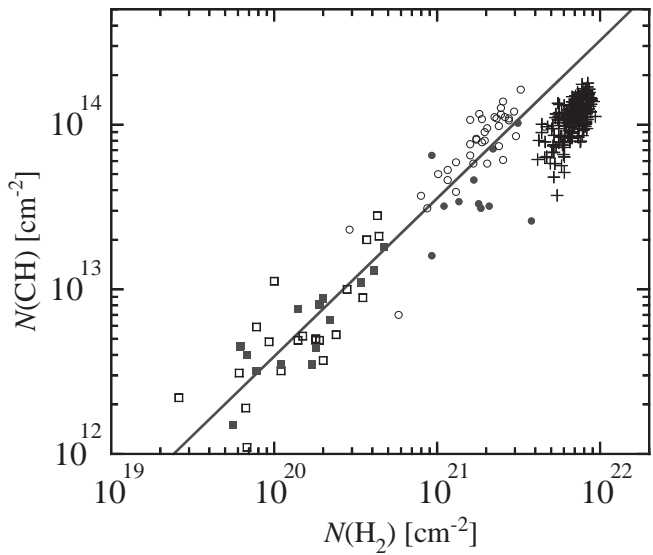}
      \caption{Correlation plot between the CH column density and the H$_2$ column density.  Cross marks denote the present data for HCL2.  Open and closed circles represent the data taken from Mattila et al. (1986) and Lang and Willson (1978), respectively.  Open and closed squares represent the data taken from Danks et al. (1984) and Federman (1982), respectively.  The solid line represents the least-square fit for all the data except for HCL2.  The solid line is almost identical with that reported by Sheffer et al. (2008).}
         \label{fig:ebv}
\end{figure}

It is known that the CH column density shows a good correlation with the H$_2$ column density in diffuse and translucent clouds with $Av<3$ (e.g. \cite{mat86, dan84, fed82, she08}).  At the same time, a hint of the turnover for higher $Av$ region is suggested in the observation toward the dark cloud L134N (\cite{mat86}).  To explore the $N$(CH)$-N$(H$_2$) relation in HCL2, we prepared the correlation diagram (Figure~\ref{fig:ebv}).  Since $N$(H$_2$) is known to be proportional to $E$(B-V), we employed the $E$(B-V) data reported by Schlegel et al. (1998) to evaluate $N$(H$_2$).  For comparison, our CH data is smoothed to the 6.1$^{\prime}$ resolution, which is the resolution of the $E$(B-V) data.  The range of $E$(B-V) of our observed positions is from 1.4 to 3.2, which corresponds to the range of  $N$(H$_2$) from $4.1 \times 10^{21}$ to $9.3 \times 10^{21}$ cm$^{-2}$.  Our CH data are not smoothly connected with the CH data taken toward the diffuse and translucent clouds.  The CH column density in the higher $N$(H$_2$) region is much lower than that expected from the correlation found in the lower $N$(H$_2$) region.

Such a break, although maybe at even lower H$_2$ column densities (of order $10^{21}$~cm$^2$, was also noted by \cite{qin10}.  These authors used the Heterodyne Instrument for the Far-Infrared (HIFI) aboard \textit{Herschel} to observe the 533 and 537 GHz fundamental spin-rotational transitions of CH in absorption toward the strong submillimeter continuum emission from the star-forming region Sgr B2(M) near the Galactic center.  In addition to absorption from the Sgr B2(M) itself, absorption is also detected over a wide LSR velocity range toward Galactic center gas as well as a series of diffuse and translucent clouds in intervening spiral arms that have H$_2$ column densities between $8 \times 10^{20}$ and $6 \times 10^{21}$~cm$^2$. The data point for Sgr B2(M) itself, which, compared to the other absorbing clouds has a much greater $N($H$_2$) of $5.4 \times 10^{23}$~cm$^2$, but a significantly (3.3--56 times) lower CH abundance, falls on the trend established by these  $N($H$_2$)$>10^{21}$~cm$^2$ absorbers.  

\subsection{CH as a tracer of cloud evolution}

The most puzzling question of our study is the origin of the broad component.  So far, the broad component (or $^{\prime}$pedestal$^{\prime}$ component) has not been recognized in the HCL2 region.  In fact, it is not seen clearly in $^{13}$CO ($J=1-0$ and $2-1$), C$^{18}$O ($J=1-0$ and $2-1$), and CI ($^3P_1 - ^3P_0$) (\cite{lan95, sch95, tat99}), probably because of overwhelming emission of the core components.  Our CH observation has clearly established its existence in HCL2, which might be an important clue for understanding the formation process of HCL2.  A broad $^{\prime}$pedestal$^{\prime}$ component of spectral lines from CH and other molecules not originating from protostellar outflows has been found in some molecular clouds and translucent clouds (e.g. \cite{mag90, mag93b, sho06}).  It has, for instance, been proposed that the $^{\prime}$pedestal$^{\prime}$ component may represent a gas flowing at the interface of two colliding clouds (e.g. \cite{ket89}).  However, it is not obvious whether this picture could be applied to HCL2.

It is known that for a cloud envelope, the velocity width increases as the cloud size increases (e.g. \cite{fal92, lar81, mye88, cas95}).  On the other hand, the velocity width is reported to be almost constant for an evolved dense core, where it is close to the thermal linewidth of H$_2$ (\cite{mye83}).  This feature was named a $^{\prime}$coherent core$^{\prime}$ by Goodman et al. (1998).  It has recently been demonstrated for the B5 cloud in Perseus that the transition from the envelope to the core is not continuous, but very sharp (\cite{pin10}).  Our observational result for CH in HCL2 seems to be consistent with the sharp transition picture in the following two points.  First, there exist the narrow and broad components in HCL2, which correspond to the core and the envelope, respectively.  The broad component is more spatially extended than the narrow component.  Second, we cannot see a gradual increase of the line width from the core to the envelope, but can only see a change in the relative peak intensity between the narrow and broad components.  This seems to reflect the sharp transition of the velocity width from the broad component to the narrow component along with the dense core formation.  Although the sharp transition is not resolved spatially with the relatively wide beam width of our observation, the transition is reflected in the line profile.  The physical mechanism of such a sharp transition is not well understood (\cite{pin10}).  However, this phenomenon might be related to the turbulence dissipation during the core formation, and understanding it is essential for a deep understanding of the physical evolution of molecular clouds.

It should be noted that Magnani et al. (1993) pointed out that the $^{\prime}$coherent$^{\prime}$ transient structure can be formed even in a turbulent medium.  In our case, it is evident that the dense cores are actually formed in the HCL2 cloud, and the narrow component traces those dense cores, as mentioned before.  Therefore, the sharp component would not be such a transient structure caused in a turbulent medium, although we need to investigate the distribution of the narrow components with higher S/N ratios and higher spatial resolution for a definitive conclusion.  A key difference between HCL2 and the translucent structures studied by Magnani et al. (1993) is that the translucent structures are not dominated by gravity, whereas HCL2 would be.

The observed increase of the narrow component contribution from the CI peak to the TMC-1 ridge is consistent with the chemical evolutionary picture.  From the CI peak to the TMC-1 ridge, dense cores are formed through the sharp transition, and the C to CO conversion proceeds through CH as an intermediate.  With the present observation of CH, the evolutionary picture of HCL2 suggested by the [CI] observations (\cite{mae99}) is further strengthened.

Our observations demonstrate that the physical and chemical evolution of molecular clouds particularly in the dense-core formation phase can well be traced by the CH $\Lambda$-type doubling transition.  This line has a low critical density ($\sim 10$~cm$^{-3}$) because of its low Einstein A coefficient ($(1-3)\times10^{-10}$~s$^{-1}$), and is optically thin.  Furthermore, CH exists both in diffuse and dense clouds with moderate abundances.  Hence, the CH line can trace a relatively wide range of physical and chemical conditions.

\subsection{Outlook}

Along with CH, the OH molecule would also trace dense core formation from diffuse clouds, which has the $\Lambda$-type doubling transitions of the $J=3/2$ state at the wavelength of 18~cm.  Harju et al. (2000) observed the OH lines toward the TMC-1 ridge with the Effelsberg 100~m telescope.  They reported that the OH distribution shows the TMC-1 ridge structure, and the intensity peak is located south of TMC-1 (CP).  The overall feature of the intensity distribution of OH is similar to that of CH in the TMC-1 ridge.  Furthermore, the OH spectral line profile seems to be composed of the narrow and broad components, as seen in the CH spectrum, although the board component is much more significant in the OH line profile.  Since OH is also an important intermediate for production of CO, it is interesting to compare the distribution of OH with that of CH.  However, the OH map by Harju et al. (2000) only covers a part of HCL2.  For such a comparison, we have recently mapped a whole region of HCL2 with the OH line with the Effelsberg 100~m telescope, which will be reported separately.

In this paper, we have studied the CH distribution in HCL2, an interesting cloud containing unique sources.  In addition to the famous carbon-chain-rich starless core, TMC-1(CP), the star-forming core, L1527, with its extraordinary richness of carbon-chain molecules resulting from warm carbon-chain chemistry (WCCC) exists in its northern part (e.g. \cite{sak07b, sak08}).  An origin of the richness of  carbon-chain molecules in these two sources might be related to the formation process of HCL2 (\cite{sak09}).  In this relation, it seems interesting to observe the Lupus-1 cloud with the CH $\Lambda$-type doubling transition lines.  The Lupus-1 cloud contains a young starless core, Lupus-1A, with abundant carbon-chain molecules like TMC-1(CP) (\cite{sak10}).  Indeed, the abundances of carbon-chain molecules are comparable to those in TMC-1(CP).  In addition, the WCCC source IRAS~15398-3359 is also associated in this region (\cite{sak09}).  A similarity of the Lupus-1 cloud and HCL2 would be related to a similarity of the formation process of dense cores, which can be explored by the large-scale distributions of the [C I], CH, and C$^{18}$O lines.

\begin{acknowledgements}
We thank Loris Magnani, the referee of this paper, for his invaluable comments and suggestions.  We are grateful to the staff of the Effelsberg 100 m telescope for excellent support.  NS thanks the Inoue Science Foundation for financial support.  This study is supported by Grant-in-Aids from the Ministry of Education,  Culture, Sports, Science, and Technologies, Japan (21224002 and 21740132).
\end{acknowledgements}



\end{document}